\documentclass[runningheads]{llncs}
\usepackage[T1]{fontenc}
\usepackage{amssymb}
\usepackage{marvosym}
\usepackage{bm}
\usepackage{amsmath}
\usepackage{hyperref}
\usepackage{diagbox}
\usepackage{subfigure}
\usepackage{xcolor}
\usepackage{multirow}
\usepackage{graphicx}
\hypersetup{colorlinks=true,
allcolors=blue
}

\begin{document}
\title{Multi-View Vertebra Localization and Identification from CT Images}
\titlerunning{Multi-View Vertebra Localization and Identification from CT Images}
%
\author{Han Wu\inst{1}
\and Jiadong Zhang\inst{1}
\and Yu Fang\inst{1}
\and Zhentao Liu\inst{1}
\and Nizhuan Wang\inst{1} 
\and \\ Zhiming Cui\inst{1}\textsuperscript{(\Letter)}
\and Dinggang Shen\inst{1,2,3}\textsuperscript{(\Letter)}
}

\authorrunning{H. Wu et al.}
%
\institute{School of Biomedical Engineering, ShanghaiTech Univerisity, Shanghai, China\\
\email{cuizm.neu.edu@gmail.com, dgshen@shanghaitech.edu.cn}\and
Shanghai United Imaging Intelligence Co. Ltd., Shanghai, China\and
Shanghai Clinical Research and Trial Center, Shanghai, China}
\maketitle   

\begin{abstract}
Accurately localizing and identifying vertebra from CT images is crucial for various clinical applications. 
However, most existing efforts are performed on 3D with cropping patch operation, suffering from the large computation costs and limited global information.
In this paper, we propose a multi-view vertebra localization and identification from CT images, converting the 3D problem into a 2D localization and identification task on different views.
Without the limitation of the 3D cropped patch, our method can learn the multi-view global information naturally.
Moreover, to better capture the anatomical structure information from different view perspectives, a multi-view contrastive learning strategy is developed to pre-train the backbone.
Additionally, we further propose a Sequence Loss to maintain the sequential structure embedded along the vertebrae.
Evaluation results demonstrate that, with only two 2D networks, our method can localize and identify vertebrae in CT images accurately, and outperforms the state-of-the-art methods consistently. Our code is available at \href{https://github.com/ShanghaiTech-IMPACT/Multi-View-Vertebra-Localization-and-Identification-from-CT-Images}{https://github.com/ShanghaiTech-IMPACT/Multi-View-Vertebra-Localization-and-Identification-from-CT-Images}.
\end{abstract}

\section{Introduction}

Automatic Localization and identification of vertebra from CT images are crucial in clinical practice, particularly for surgical planning, pathological diagnosis, and post-operative evaluation\cite{burns2016automated,knez2016computer,kumar2014robotic}. 
However, the process is challenging due to the significant shape variations of vertebrae with different categories, such as lumbar and thoracic, and also the close shape resemblance of neighboring vertebrae.
Apart from these intrinsic challenges, the arbitrary field-of-view (FOV) of different CT scans and the presence of metal implant artifacts also introduce additional difficulties to this task.

With the advance of deep learning, many methods are devoted to tackling these challenges. 
For example, Lessmann et al. \cite{lessmann2019iterative} employed a one-stage segmentation method to segment vertebrae with different labels for localization and identification.
It is intuitive but usually involves many segmentation artifacts.
Building upon this method, Masuzawa et al.\cite{masuzawa2020automatic} proposed an instance memory module to capture the neighboring information, but the long-term sequential information is not well studied.
Recently, two or multi-stage methods\cite{cheng2021automatic,cuivertnet,meng2021vertebrae,payer2020coarse,qin2021vertebrae}, that first localize the vertebra and further classify the detected vertebra patches, are proposed to achieve the state-of-the-art performance. 
And some additional modules, such as attention mechanism\cite{cuivertnet}, graph optimization\cite{meng2021vertebrae}, and LSTM\cite{qin2021vertebrae}, are integrated to capture the sequential information of adjacent vertebrae. 
However, all these methods are performed on 3D patches, where the global information of the CT scan is destroyed and cannot be well-captured.
Moreover, due to the lack of pre-trained models in 3D medical imaging, networks trained from scratch using a small dataset often lead to severe overfitting problems with inferior performance.

In this paper, to tackle the aforementioned challenges, we present a novel framework that converts the 3D vertebra labeling problem into a multi-view 2D vertebra localization and identification task. 
Without the 3D patch limitation, our network can learn 2D global information naturally from different view perspectives, as well as leverage the pre-trained models from ImageNet\cite{deng2009imagenet}.
Specifically, given a 3D CT image, we first generate multi-view 2D Digitally Reconstructed Radiograph (DRR) projection images.
Then, a multi-view contrastive learning strategy is designed to further pre-train the network on this specific task. 
For vertebra localization, we predict the centroid of each vertebra in all DRR images and map the 2D detected centroids of different views back into the 3D CT scan using a least-squares algorithm. 
As for vertebra identification, we formulate it as a 2D segmentation task that generates vertebra labels around vertebra centroids. 
Particularly, a Sequence Loss, based on dynamic programming, is introduced to maintain the sequential information along the spine vertebrae in the training stage, which also serves as a weight to vote the multi-view 2D identification results into the 3D CT image for more reliable results.
Our proposed method is validated on a public challenging dataset\cite{VerSe} and achieved the state-of-the-art performance both in vertebra localization and identification.
Moreover, more evaluation results on a large-scale in-house dataset collected in real-world clinics (with 500 CT images) are provided in the supplementary materials, further demonstrating the effectiveness and robustness of our framework.

\section{Methodology}

\begin{figure*}[t]
    \centering
    \includegraphics[width=\textwidth]{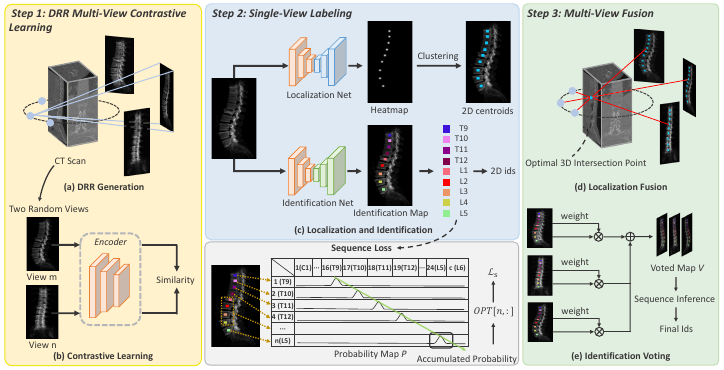}
    \caption{An overview of the proposed method, including (a) DRR Generation, (b) DRR multi-view contrastive learning, (c) single-view vertebra localization and identification, (d) multi-view localization fusion, and (e) multi-view identification voting. The implementation of Sequence Loss is also illustrated.}
    \label{fig:pipeline}
\end{figure*}

An overview of our proposed method for vertebra localization and identification using multi-view DRR from CT scans is shown in Fig. \ref{fig:pipeline}, which mainly consists of three steps. Step 1 is to generate DRR images, followed by a multi-view contrastive learning strategy to pre-train the backbone. Step 2 aims to finish 2D single-view vertebra localization and identification, and step 3 is to map the 2D results back to 3D with a multi-view fusion strategy. We will elaborate our framework in this section.

\subsection{DRR Multi-View Contrastive Learning}
\subsubsection{DRR Generation.}
To accurately localize and identify each vertebra in CT images, we convert the 3D task into 2D, where global information can be naturally captured from different views, avoiding the large computation of 3D models.
To achieve this, DRR (Digitally Reconstructed Radiograph) technique, a simulation procedure for generating a radiograph similar to conventional X-ray image, is performed by projecting a CT image onto a virtual detector plane with a virtual X-ray source. In this way, we can generate $K$ DRR projection images of a CT image for every $360/K$ degree. The 3D labeling problem can then be formulated as a multi-view localization and identification task in a 2D manner. Specifically, the 2D ground-truth are generated by projecting the 3D centroids and labels onto the 2D image following the DRR projection settings. 

\subsubsection{DRR Multi-View Contrastive Learning.} 
After DRR generation, our goal is to localize and identify the vertebra on DRR images.
However, as the dataset for the vertebra task is relatively small due to time-consuming manual annotation, we design a new multi-view contrastive learning strategy to better learn the vertebrae representation from various views.
Unlike previous contrastive learning methods, where the pretext is learned from numerous augmented negative and positive samples\cite{SimCLR,simsiam,BYOL,MoCoV1}, e.g., random crop, image flip, rotation and resize, the multi-view DRR images generated from the same CT image share consistent anatomical information, which are natural positive samples.
Based on this insight, we pre-train our network backbone using the Simsiam \cite{simsiam} approach to encode two random views from the same CT image as a key and query, as shown in Fig. \ref{fig:pipeline} (b), in the aims of learning the invariant vertebrae representation from different views.


\subsection{Single-View Vertebra Localization}
With multi-view DRR images, the 3D vertebra localization problem is converted into a 2D vertebra centroid detection task, followed by a multi-view fusion strategy (as introduced in Sec. \ref{sec:multu-view}) that transforms the 2D results to 3D.
To achieve this, we utilize the commonly-used heatmap regression strategy for 2D vertebra centroid detection. 
Specifically, for each vertebra in a DRR image, our model is trained to learn the contextual heatmap defined on the ground-truth 2D centroid using a Gaussian kernel. 
During inference, we apply a fast peak search clustering method \cite{rodriguez2014clustering} to localize the density peaks on the regressed heatmap as the predicted centroid. 
Benefiting from the pre-trained models from multi-view contrastive learning, our method can capture more representative features from different views.
Further, compared to existing 3D methods, our approach performs vertebra localization on several DRR images with a fusion strategy, making it more robust to the situation of missing detection in certain views.

\subsection{Single-View Vertebra Identification}
After the vertebrae localization, we further predict the label of each vertebra using an identification network on multi-view DRR images. 
Unlike other 3D methods that require cropping vertebra patches for classification, our identification network performs on 2D, allowing us to feed the entire DRR image into the network, which can naturally capture the global information. 
Specifically, we use a segmentation model to predict the vertebra labels around the detected vertebra centroids, i.e., a $22 mm \times 22 mm$ square centered at the centroid. 
During the inference of single-view, we analyze the pixel-wise labels in each square and identify the corresponding vertebra with the majority number of labels.

\subsubsection{Sequence Loss.}

In the identification task, we observe that the vertebra labels are always in a monotonically increasing order along the spine, which implies the presence of sequential information. 
To better exploit this property and enhance our model to capture such sequential information, we propose a Sequence Loss as an additional network supervision, ensuring the probability distribution along the spine follows a good sequential order.
Specifically, as shown in Fig. \ref{fig:pipeline}, we compute a probability map $P\in \mathbb{R}^{n \times c}$ for each DRR image by averaging the predicted pixel-wise possibilities in each square around the vertebra centroid from the identification network. Here, $n$ is the number of vertebrae contained in this DRR image, and $c$ indicates the number of vertebra categories (i.e., from C1 to L6).  
Due to the sequential nature of the vertebra identification problem, the optimal distribution of $P$ is that the index of the largest probability in each row is in ascending order (green line in Fig. \ref{fig:pipeline}).
To formalize this notion, we compute the largest accumulated probability in ascending order, starting from each category in the first row and ending at the last row, using dynamic programming. 
The higher accumulated probability, the better sequential structure presented by this distribution. 
We set this accumulated probability as target profit, and aim to maximize it to enable our model to better capture the sequential structure in this DRR image. 
The optimal solution ($OPT$) based on the dynamic programming algorithm is as:

\begin{equation}
\label{OPT}
\begin{split}
		OPT[i,j]=&
		\begin{cases}
			P[i,j] \quad  &if \, j = 1\, or\, i = 1 \\
			OPT[i-1,j-1] + D \quad &otherwise
		\end{cases}\\
  D &= \max(\alpha P[i,j-1], \beta P[i,j] , \alpha P[i,j+1]),
\end{split}
\end{equation}
where $i\in[1, n]$ and $j\in[1, c]$. Here, $\alpha$ and $\beta$ are two parameters that are designed to alleviate the influence of wrong-identified vertebra. Sequence Loss (${\cal L}_{s}$) is then defined as:
\begin{equation}
\label{Sequence_loss}
\begin{split}
    {\cal L}_{s} = 1 - \frac{max(OPT[n, :])}{\beta n} .\\
\end{split}
\end{equation}

The overall loss function ${\cal L}_{id}$ for our identification network is:
\begin{equation}
    {\cal L}_{id} = {\cal L}_{ce} + \gamma {\cal L}_{s} ,
\end{equation}
where ${\cal L}_{ce}$ and ${\cal L}_{s}$ refer to the Cross-Entropy loss and Sequence Loss, respectively. $\gamma$ is a parameter to control the relative weights of the two losses.

\subsection{Multi-View Fusion}
\subsubsection{Localization Multi-View Fusion.}
\label{sec:multu-view}
After locating all the vertebrae in each DRR image, we fuse and map the 2D centroids back to 3D space by a least-squares algorithm, as illustrated in Fig. \ref{fig:pipeline} (d). 
For a vertebra located in $K$ views, we can track $K$ target lines from the source points in DRR technique to the detected centroid on the DRR images. 
Ideally, the $K$ lines should intersect at a unique point in the 3D space, but due to localization errors, this is always unachievable in practice. 
Hence, instead of finding a unique intersection point, we employ the least-squares algorithm to minimize the sum of perpendicular distances from the optimal intersection point to all the $K$ lines, given by:
\begin{equation}
    D(\bm{p};\bm{A},\bm{N}) = \sum_{k=1}^{K}D(\bm{p};\bm{a}_k,\bm{n}_k)=\sum_{k=1}^{K}(\bm{a}_k-\bm{p})^T(\bm{I}-\bm{n}_{k}\bm{n}_{k}^T)(\bm{a}_k-\bm{p}) ,
\end{equation}
where $\bm{p}$ denotes the 3D coordinate of the optimal intersection point, $\bm{a}_k$ and $\bm{n}_k$ represent the point on the $k_{th}$ target line and the corresponding direction vector. 
By taking derivatives with respect to $\bm{p}$, we get a linear equation of $\bm{p}$ as shown in Eq. (\ref{intersection_point}), where the optimal intersection point can be obtained by achieving the minimum distance to the $K$ lines.
\begin{equation}
\label{intersection_point}
\centering
    \begin{split}
    \frac{\partial D}{\partial \bm{p}} = \sum_{k=1}^{K} - 2(\bm{I}-\bm{n}_k\bm{n}_k^{T})(\bm{a}_k-\bm{p}) = 0 \Rightarrow \bm{Sp} = \bm{q} ,\\
    \bm{S} = \sum_{k=1}^{K}(\bm{I}-\bm{n}_{k}\bm{n}_{k}^T) , \; \bm{q} = \sum_{k=1}^{K}(\bm{I}-\bm{n}_{k}\bm{n}_{k}^T)\bm{a}_k .\\
    \end{split}
\end{equation}

\subsubsection{Identification Multi-View Voting.}
The Sequence Loss evaluates the quality of the predicted vertebra labels in terms of their sequential property. 
During inference, we further use this Sequence Loss of each view as weights to fuse the probability maps obtained from different views. We obtain the final voted identification map $V$ of $K$ views as:
\begin{equation}
    V = \sum_{k=1}^{K}W_{k} P_{k},\,\, W_k = \frac{(1-{\cal L}_{s}^k)}{\sum_{a=1}^{K}(1-{\cal L}_{s}^a)}.
\end{equation}

For each vertebra, the naive solution for obtaining vertebra labels is to extract the largest probability from each row in voted identification map $V$. 
Despite the promising performance of the identification network, we still find some erroneous predictions. 
To address this issue, we leverage the dynamic programming (described in Eq. (\ref{OPT})) again to correct the predicted vertebra labels in this voted identification map $V$.
Specifically, we identify the index of the largest accumulated probability in the last row as the last vertebra category and utilize it as a reference to correct any inconsistencies in the prediction.

\section{Experiments and Results}
\subsection{Dataset and Evaluation Metric}
We extensively evaluate our method on the publicly available MICCAI VerSe19 Challenge dataset \cite{VerSe}, which consists of 160 spinal CT with ground truth annotations.
Specifically, following the public challenge settings, we utilize 80 scans for training, 40 scans for testing, and 40 scans as hidden data. 
To evaluate the performance of our method, we use the mean localization error (L-Error) and identification rate (Id-Rate) as the evaluation metrics, which are also adopted in the challenge.
The L-Error is calculated as the average Euclidean distance between the ground-truth and predicted vertebral centers. 
The Id-Rate is defined as the ratio of correctly identified vertebrae to the total number of vertebrae.

\subsection{Implementation Details}
All CT scans are resampled to an isotropic resolution of 1 mm. 
For DRR Multi-View Contrastive Learning, we use ResNet50 as encoder and apply the SGD optimizer with an initial learning rate of 0.0125, which follows the cosine decay schedule.
The weight decay, SGD momentum, batch size and loss function are set to 0.0001, 0.9, 64, and cosine similarity respectively.
We employ U-Net for both the localization and identification networks, using the pre-trained ResNet50 from our contrastive learning as backbone. 
Adam optimizer is set with an initial learning rate of 0.001, which is divided by 10 every 4000 iterations. 
Both networks are trained for 15k iterations. 
We empirically set $\alpha = 0.1$, $\beta=0.8$, $\gamma = 1$. 
All methods were implemented in Python using PyTorch framework and trained on an Nvidia Tesla A100 GPU with 40GB memory.

\subsection{Comparison with SOTA Methods }

\begin{table}[t]
\centering
\caption{Results on the VerSe19 challenge dataset.}
\label{tab:verse19result}
\begin{tabular}{l|llll}
\hline
Method        & \multicolumn{2}{l|}{Test Dataset}              & \multicolumn{2}{l}{Hidden Dataset}                                                             \\ \cline{2-5} 
              & \multicolumn{1}{l|}{Id-Rate(\%)} & \multicolumn{1}{l|}{L-Error(mm)} & \multicolumn{1}{l|}{Id-Rate(\%)} & L-Error(mm) \\ \hline
Payer C.\cite{VerSe}      & \multicolumn{1}{l|}{95.65}       & \multicolumn{1}{l|}{4.27}        & \multicolumn{1}{l|}{94.25}       & 4.80        \\
Lessmann N.\cite{VerSe}   & \multicolumn{1}{l|}{89.86}       & \multicolumn{1}{l|}{14.12}       & \multicolumn{1}{l|}{90.42}       & 7.04        \\
Chen M. \cite{VerSe}      & \multicolumn{1}{l|}{96.94}       & \multicolumn{1}{l|}{4.43}        & \multicolumn{1}{l|}{86.73}       & 7.13        \\
Sekuboyina A.\cite{sekuboyina2018btrfly} & \multicolumn{1}{l|}{89.97}       & \multicolumn{1}{l|}{5.17}        & \multicolumn{1}{l|}{87.66}       & 6.56        \\
Ours          & \multicolumn{1}{l|}{\underline{\textbf{98.12}}}       & \multicolumn{1}{l|}{\underline{\textbf{1.79}}}        & \multicolumn{1}{l|}{\underline{\textbf{96.45}}}       & \underline{\textbf{2.17}}        \\ \hline
\end{tabular}
\end{table}

We train our method on 70 CT images and tune the hyperparameter on the rest 10 CT images from the training data. We then evaluate it on both testing and hidden datasets, following the same setting as the challenge.
In the comparison, our method is compared with four methods which are the first four positions in the benchmark of this challenge\cite{VerSe}. 
The experimental results are presented in Table \ref{tab:verse19result}. 
Our method achieves Id-Rate of 98.12\% and L-Error of 1.79 mm on the test dataset, and Id-Rate of 96.45\% and L-Error of 2.17 mm on the hidden dataset, which achieves the leading performance both in localization and identification tasks with just two 2D networks.
Compared to these methods performed on 3D with random cropping or patch-wise method (Payer C.\cite{VerSe}, Lessmann N.\cite{VerSe} and Chen M.\cite{VerSe}), our 2D strategy can capture more reliable global and sequential information in all 2D projection images which can improve the labeling performance, especially the localization error. Compared to those using 2D MIP(Sekuboyina A.\cite{sekuboyina2018btrfly}), our DRR multi-view projection and fusion strategy can provide superior performance by analyzing more views and introducing the geometry information carried by varied DRR projections naturally.

\subsection{Ablation Study}

\begin{table}[t]
\centering
\caption{Ablation study results of key components.}
\label{tab:alation_study}
\begin{tabular}{cccc|ll}
\hline
\multirow{2}{*}{\,Baseline\,}           & \multirow{2}{*}{\,Pre-train\,}          & \multirow{2}{*}{\,Sequence Loss\,}          & \multirow{2}{*}{\,Voting\,} & \multicolumn{2}{l}{Id-Rate(\%)}                    \\ \cline{5-6} 
                                    &                                     &                                         &                         & \multicolumn{1}{l|}{Test Dataset} & Hidden dataset \\ \hline
$\checkmark$                       &                                     &                                          &                         & \multicolumn{1}{l|}{84.00}        & 83.45          \\
$\checkmark$                       & $\checkmark$                        &                                          &                         & \multicolumn{1}{l|}{85.58}        & 86.52          \\
$\checkmark$                       & $\checkmark$                        & $\checkmark$                            &                         & \multicolumn{1}{l|}{89.41}        & 90.54          \\
$\checkmark$                       & $\checkmark$                        & $\checkmark$                            & $\checkmark$                     & \multicolumn{1}{l|}{\underline{\textbf{98.12}}}        & \underline{\textbf{96.45}}          \\ \hline
\end{tabular}
\end{table}

\begin{figure}[htb]
\centering
    \subfigure[]{\includegraphics[width=0.32\textwidth, height=7cm]{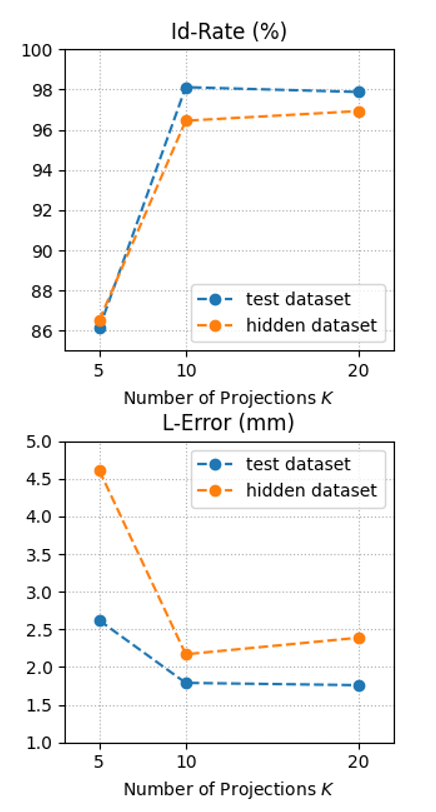}}
    \subfigure[]{\includegraphics[width=0.55\textwidth, height=7cm]{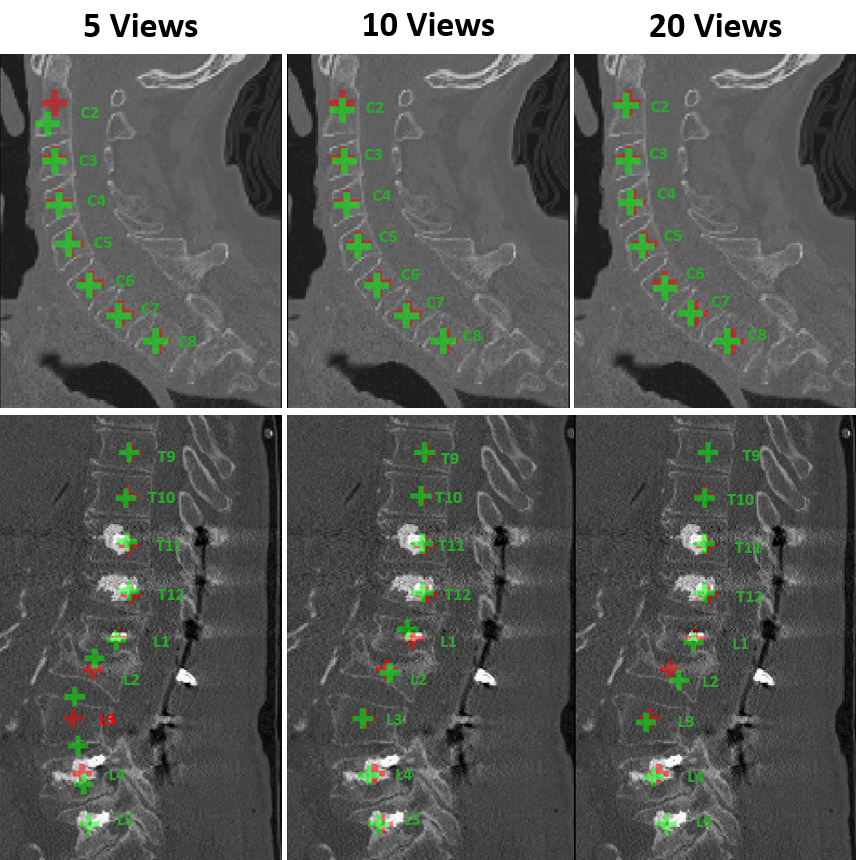}}
    \caption{(a) The Id-Rate and L-Error of different $K$. (b) Comparison between different $K$ from the final predicted CT scan on limited FOV and metal artifacts cases (red for ground truth and green for predictions).}
    \label{fig:K_ablation}
\end{figure}

\subsubsection{Ablation Study of Key Components.}
We conduct an ablation study on the VerSe19 dataset to demonstrate the effectiveness of each component.
As presented in Table \ref{tab:alation_study}, we build the basic network for the vertebra localization and identification with a bagging strategy, where for each vertebra, we opt for the ID that is predicted by the majority of views, when not using weighted voting, and $K=10$, denoted as Baseline. 
Pre-train, Sequence Loss, and voting in Table \ref{tab:alation_study} represent the addition of the multi-view contrastive learning, Sequence Loss, and multi-view voting one by one. Pre-trained from ImageNet is used when not utilizing our contrastive learning pre-trained parameters. 
Specifically, the Baseline achieves Id-Rate of 84.00\% and 83.54\% on two datasets. 
With the contrastive learning pre-trained parameters, we achieve 1.88\% and 2.98\% improvements over the ImageNet pre-trained, respectively. This shows the pre-trained parameters of the backbone obtained from our contrastive learning can effectively facilitate the network to learn more discriminative features for identification than the model learning from scratch.
Sequence Loss provides extra supervision for sequential information, and results in 3.53\% and 4.02\% increase, illustrating the significance of capturing the sequential information in the identification task.
Finally, multi-view weighted voting yields the best results with 98.12\% and 96.45\% on the two datasets, indicating the robustness of our multi-view voting when the identification errors occurred in a small number of DRR images can be corrected by other DRR prediction results.

\subsubsection{Ablation Study of Projection Number.} 
We also conduct an ablation study on the same dataset to further evaluate the impact of the projection number $K$. 
The results are presented in Fig. \ref{fig:K_ablation}, indicating a clear trend of performance improvements as the number of projections $K$ increases from 5 to 10. 
However, when $K$ increases to 20, the performance is just comparable to that of 10. 
We analyze that using too few views may result in inadequate and unreliable anatomical structure representation, leading to unsatisfactory results. 
On the other hand, too many views may provide redundant information, resulting in comparable results but with higher computation cost. Therefore, $K$ is set to 10 as a trade-off between accuracy and efficiency.

\section{Conclusion}
In this paper, we propose a novel multi-view method for vertebra localization and identification in CT images.
The 3D labeling problem is converted into a multi-view 2D localization and identification task, followed by a fusion strategy.
In particular, we propose a multi-view contrastive learning strategy to better learn the invariant anatomical structure information from different views.
And a Sequence Loss is further introduced to enhance the framework to better capture sequential structure embedded in vertebrae both in training and inference. Evaluation results on a public dataset demonstrate the advantage of our method.

%
%
%
\bibliographystyle{splncs04}
\bibliography{references.bib}

\end{document}


\title{Multi-View Vertebra Localization and Identification from CT Images \\ Supplementary Material}
%
%
\author{Paper ID: 534}
\authorrunning{Paper ID: 534}
%
\institute{}
\maketitle              

\begin{table}
\centering
\caption{Evaluation results on a large-scale in-house dataset collected from the practical clinics with 500 CT scans divided into 300 for training, 100 for testing, and 100 for validation. We train the model on the training dataset, and further evaluate it on the test and validation dataset with $K$ set to 10.}
\begin{tabular}{l|ll|ll}
\hline
     & \multicolumn{2}{l|}{Test dataset}              & \multicolumn{2}{l}{Validation dataset}         \\ \cline{2-5} 
     & \multicolumn{1}{l|}{Id-Rate(\%)} & L-Error(mm) & \multicolumn{1}{l|}{Id-Rate(\%)} & L-Error(mm) \\ \hline
Cer. & \multicolumn{1}{l|}{99.67}       & 1.31       & \multicolumn{1}{l|}{99.55}       & 1.51       \\
Tho. & \multicolumn{1}{l|}{98.24}       & 1.34       & \multicolumn{1}{l|}{99.00}       & 1.48       \\
Lum. & \multicolumn{1}{l|}{99.31}       & 1.35       & \multicolumn{1}{l|}{99.54}       & 1.50       \\ 
All  & \multicolumn{1}{l|}{98.62}       & 1.34       & \multicolumn{1}{l|}{99.04}       & 1.49       \\ \hline
\end{tabular}
\end{table}

\begin{figure*}[h]
    \centering
    \includegraphics[width=\textwidth]{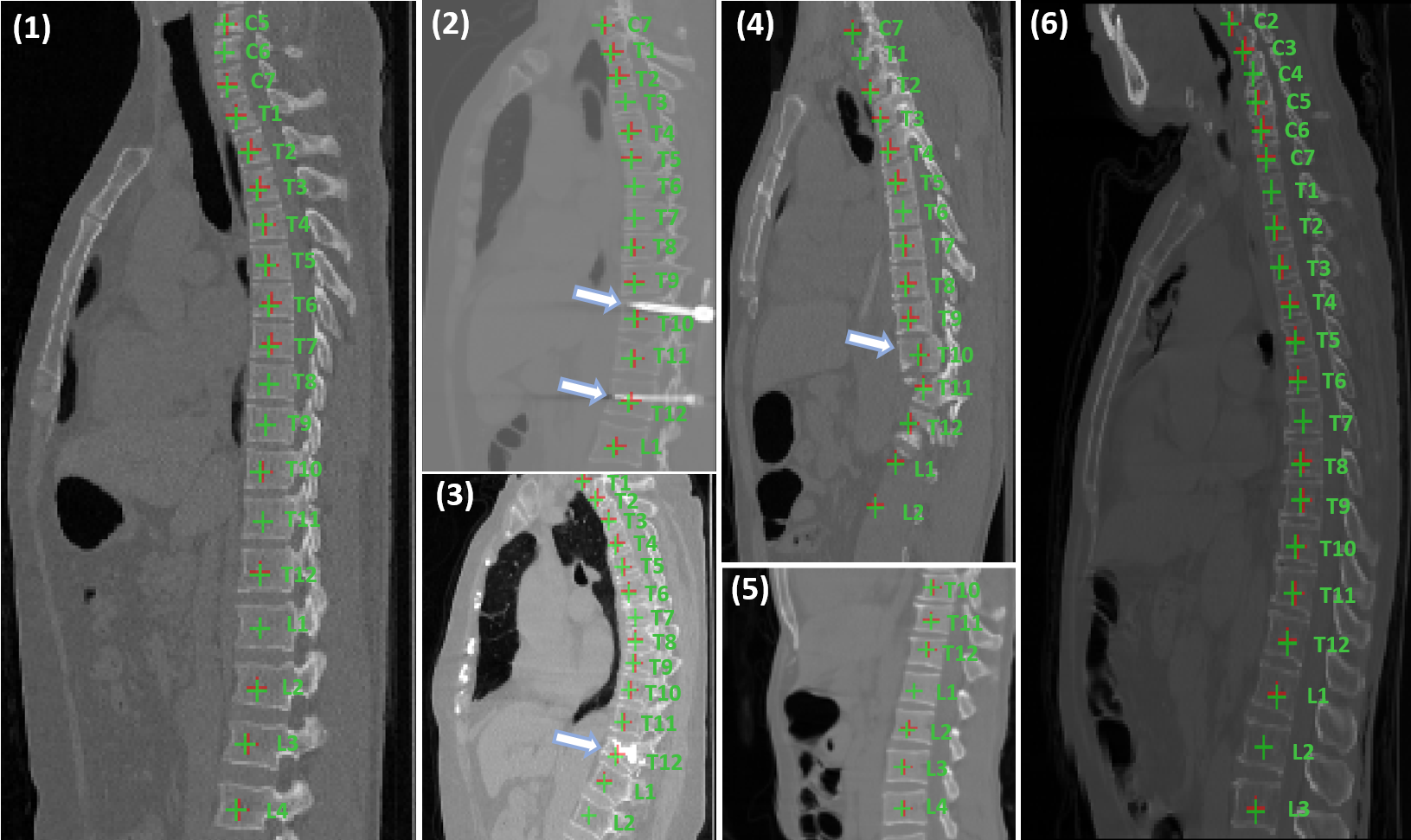}
    \caption{Qualitative results on typical challenging cases: large field of view(1, 6), metal artifacts (2, 3), pathological spines (4), and limited field of view (5).(Green for prediction and red for ground truth.)}
    \label{fig:pipeline}
\end{figure*}